\documentstyle[aps,prl,twocolumn, epsf]{revtex}

\def\be{\begin{equation}}
\def\ee{\end{equation}}
\def\bea{\begin{eqnarray}}
\def\eea{\end{eqnarray}}
\def\bma{\begin{mathletters}}
\def\ema{\end{mathletters}}
\newcommand{\bra}[1]{\mbox{$\langle #1 |$}}
\newcommand{\ket}[1]{\mbox{$| #1 \rangle$}}

\newcommand{\proj}[1]{\ket{#1}\!\bra{#1}}

\begin{document}

\draft

\title{Optimal estimation of two-qubit pure-state entanglement}

\author{Antonio Ac\'{\i}n, Rolf Tarrach and Guifr\'e Vidal}

\address{
%e-mail: acin@ecm.ub.es\\
 Departament d'Estructura i Constituents de la Mat\`eria,
 Universitat de Barcelona, Diagonal 647, E-08028 Barcelona, Spain.
}

\date{\today}

\maketitle

\begin{abstract}
We present optimal measuring strategies for the estimation of the entanglement of unknown two-qubit pure states and of the degree of mixing of unknown single-qubit mixed states, of which $N$ identical copies are available. The most general measuring strategies are considered in both situations, to conclude in the first case that a local, although collective, measurement suffices to estimate entanglement, a non-local property, optimally.
\end{abstract}

\pacs{PACS Nos. 03.67.-a, 03.65.Bz}

\bigskip

\section{Introduction}

%
% introduction
%

Plenty of work has been performed in recent years on optimal quantum measurements, i.e. on measurements which provide the maximum possible information about an unknown quantum mechanical pure \cite{HOL,MP,DBE,LPT,ALP} or mixed \cite{VLPT} state, of which $N$ identical copies are available. These works are focussed mainly on the determination of the unknown state as a whole, and consequently any of its properties is also estimated, although maybe not in an optimal way.

On the other hand recent developments on the field of quantum information theory have stressed the importance of the quantum correlations --or entanglement-- displayed by some states of composite systems. In the simplest of such composite systems, the two-qubit case, all non-local properties of pure states depend upon only one single parameter. Such non-local parameter is the only relevant quantity invariant under local unitary transformations on each qubit and plays a central role in the quantification and optimal manipulation of entanglement \cite{BEN,LOPOP,NIE,VID,JON}.

%
% What we do
%

 In this work we analyze and solve the problem of optimally estimating the entanglement of an unknown pure state of two qubits. This problem has been independently addressed also by Sancho and Huelga in a recent work \cite{Susana}, where only a restricted class of measuring strategies is considered. Here, on the contrary, we will consider most general quantum measurements on $N$ identical copies of the state. Their quality will be assessed through the gain of information they provide about the non-local parameter of the state. After presenting and proving the solution we will conclude that the optimal measuring strategies so defined are not equivalent to the ones used to fully reconstruct the unknown state. As a matter of fact, {\em all} information about some relative phase of the unknown state turns out to be irreversibly erased as the entanglement is estimated.

 Estimation of the degree of mixing of an unknown mixed state is a different but very much related topic that we shall also consider here. For the single-qubit case the amount of mixing is specified again by just one parameter, the modulus of the corresponding Bloch vector, whereas in order to completely specifying the state two more parameters, namely the direction of the Bloch vector, are also required. We shall show that in this case the optimal measuring strategy on any number $N$ of qubits prepared in the same mixed state can be made compatible with the optimal estimation of the direction of its Bloch vector.

 Finally, we will show that a possible way of optimally determining the entanglement of an unknown, two-qubit pure state consists precisely in estimating, also optimally, the degree of mixture of any of its two reduced density matrices. Therefore, it turns out in this simple bipartite case that the optimal estimation of a non-local parameter can be done through a local measurement.

%
% structure of the paper
%

 The paper is structured as follows. Section \ref{sec:background} is devoted to background material. We introduce a convenient parameterization of two-qubit pure states and consider their isotropic distribution. We also review some basic aspects on parameter estimation and on quantum measurements. In Section \ref{sec:result}  we pose the problem of entanglement estimation on firmer grounds and announce the main result of this paper: its optimal performance. Section \ref{sec:computation}, rather technical and that could well be skipped in a first reading, is devoted to the computation of some effective density matrix $\rho^{(N)}(b)$, an object which plays a central role in deriving the optimal strategy for estimating entanglement. In Section \ref{sec:examples} the $N=1,2,3$ cases are presented in more detail in order to illustrate the general case. Optimal estimation of the degree of mixing is discussed and solved in Section \ref{sec:mix}, and finally Section \ref{sec:conclusions} contains a discussion relating estimation of both entanglement and mixing, and some concluding remarks.

\section{Preliminaries}
\label{sec:background}

 We will consider here a two-party scenario. Alice and Bob will share the $N$ copies of a completely unknown two-qubit pure state $\ket{\psi}$, and their aim will be to obtain as much information as possible about its entanglement. The sense in which the state is {\em unknown}, the mechanisms for {\em extracting} information from the system and the scheme for {\em evaluating} the extracted information will be briefly reviewed in what follows.

\subsection{Homogeneous distribution.}

%
% parameterization and homogeneous distribution
%

 All that is initially known about the state of each pair of qubits is that it is pure. This corresponds to the unbiased distribution on the Hilbert space ${\mathcal H}_4={\mathcal H}_2\otimes{\mathcal H}_2$ of two qubits, that is, to the only probability distribution invariant under arbitrary unitary transformations on ${\mathcal H}_4$. It is convenient to express the unknown state $|\psi\rangle\in{\mathcal H}_2\otimes{\mathcal H}_2$, which depends on six parameters, in its Schmidt-like decomposition

\begin{equation}
  |\psi\rangle=\sqrt{\frac{1+b}{2}}|\hat{a}\rangle|\hat{b}\rangle+
    \sqrt{\frac{1-b}{2}}e^{i\alpha}|-\hat{a}\rangle|-\hat{b}\rangle ,
\label{Schmidt}
\end{equation}
where the phase $e^{i\alpha}$, which is usually absorbed by one of the
kets it goes with, has been left explicit. The non-local parameter $b~\in~[0,1]$ characterizes the entanglement of $\ket{\psi}$. Only for $b=1$ is $\ket{\psi}$ a product state $\ket{\hat{a}}\otimes\ket{\hat{b}}$, and thus unentangled. For $b<1$ the state contains quantum correlations, $b=0$ corresponding to a maximally entangled state. Recall that this parameter is the modulus of the Bloch vector of the reduced density matrix $\rho_A$ on Alice's side,
\begin{equation}
  \rho_A\equiv tr_B|\psi\rangle\langle\psi|=\frac{1+b}{2}|\hat{a}\rangle
    \langle\hat{a}|+\frac{1-b}{2}|-\hat{a}\rangle\langle-\hat{a}| ,
\label{redmatr}
\end{equation}
and equivalently for $\rho_B$. The other four parameters correspond to
the two directions $\hat{a}$ and $\hat{b}$ of the Bloch vectors of $\rho_A$ and $\rho_B$. Then, the unbiased distribution of pure states corresponds \cite{HA} to the isotropic distribution of $\hat{a}$ in $S^2$, $\hat{b}$ in $S^2$, $\alpha$ in $S^1$ and the quadratic distribution of
$b$ in [0,1],
\begin{equation}
  \int\limits_{S^2}\frac{d\hat{a}}{4\pi}\int\limits_{S^2}\frac{d\hat{b}}
    {4\pi}\int\limits_{S^1}\frac{d\alpha}{2\pi}\int_0^1 db\thinspace
    3b^2=1 .
\label{norm}
\end{equation}

\subsection{General measurements and information gain.}

%
% general measurements and average Kullback
%

The parties are thus provided with $N$ copies of a pure state $\ket{\psi}$ as in Eq. (\ref{Schmidt}), i.e. with the state $\ket{\psi}^{\otimes N}$, and our aim is to construct the most informative measurement on the collective, $2N$-qubit system for the estimation of the parameter $b$. The optimality criterion to be used is based on the Kullback or mutual information $K[f',f]$ \cite{KU}, a functional of two probability distributions $f'$ and $f$ that is interpreted as the gain of information in replacing the latter distribution with the former one\cite{HOB}. In our case, for instance, the prior, unbiased density function for the parameter $b$ is given by (\ref{norm}), so we have
$f(b)=3b^2$. A generic measurement, allowing for the most general manipulation of the system, is represented by a resolution of the identity by means of a set of positive operators,
\begin{equation}
  \sum_kM^{(k)}=I .
\label{POVM}
\end{equation}
After the above positive operator valued measurement (POVM) has been performed, giving the outcome $k$ with probability $\mbox{tr}(M^{(k)}\rho^{\otimes N})$, where $\rho=|\psi\rangle\langle\psi|$, we
compute the posterior density function for $b$, $f(b\vert k)$, through the Bayes formula
\begin{equation}
  f_k(b)\equiv f(b\vert k)=\frac{p(k\vert b)f(b)}{p(k)},
\label{Bayes}
\end{equation}
where $p(k)$ is given by
\begin{equation}
  p(k)=\int_0^1 dbf(b)p(k\vert b) ,
\label{probk}
\end{equation}
and the conditional probability of getting outcome $k$ when the state's non-local parameter has value $b$, $p(k\vert b)$, will be shown later.
The gain of information resulting from obtaining the outcome $k$
after the measurement is quantified by the Kullback information corresponding to the prior and posterior probability density functions
\begin{equation}
  K[f_k, f]=\int dbf(b\vert k)\ln\left(\frac{f(b\vert k)}{f(b)}
    \right) .
\label{Kullback}
\end{equation}
This expression has to be averaged over all the possible outcomes of the
measurement, so that the expected gain of information reads
\begin{equation}
  \bar{K}[f_k ,f]=\sum_kp(k)K[f_k, f] ,
\label{avKul}
\end{equation}
and using (\ref{Bayes}) this expression can be written as
\begin{equation}
  \bar{K}[f_k ,f]=\sum_k\int dbf(b)p(k\vert b)\ln\left(\frac
    {p(k\vert b)}{p(k)}\right) .
\label{avKul2}
\end{equation}

%
% choice of parameter
%

 Let us notice here that the value of $K[f_k, f]$ in Eq. (\ref{Kullback}) would remain unchanged if we decided to characterize the entanglement of $\ket{\psi}$ by another parameter $b´=h(b)$ (where $h(b)$ is any bijective function of the original parameter $b$). Consequently, the gain of information we compute for $b$ also applies to any of the measures of entanglement so far proposed, such as the entanglement of formation \cite{BEN}
\be
-\sqrt{\frac{1+b}{2}}\log_2 \sqrt{\frac{1+b}{2}} -\sqrt{\frac{1-b}{2}}\log_2 \sqrt{\frac{1-b}{2}}
\ee  
for the asymptotic regime, or the monotone \cite{VID}
\be
\sqrt{\frac{1-b}{2}}
\ee
for the single-copy case.

\section{Optimal measurements for entanglement estimation}
\label{sec:result}

%
% plan
%

 We are looking for a measurement of the form (\ref{POVM}) such that the expected gain of information (\ref{avKul2}) is maximized. We will present and explain here and in Section \ref{sec:examples} such optimal measurements, whereas their explicit construction is mainly contained in Section \ref{sec:computation}.

\subsection{Local and global strategies}
\label{subsec:types_measurement}
%
% types of measurements
%

 Before we proceed we comment on four classes of measurements Alice and Bob may consider in order to learn about $b$ \cite{Susana}: 
\begin{itemize}

\item {\em local} measurements on only, say, Alice's side, i.e. on the $N$ qubits supporting the local state $\rho_A^{\otimes N}$, would be the most restrictive class of the hierarchy;
\item {\em uncorrelated bilocal} --i.e. each party measuring on their local $N$-qubit part independently-- and
\item {\em classically correlated bilocal} --that is, with classical communication between Alice and Bob-- measurements are two intermediate types of strategies; finally,
\item {\em global} measurements on the $2N$ qubits constitute the most general case.
\end{itemize}
 Global measurements are in principle the most informative ones. But as the parameter $b$ which quantifies the entanglement of $|\psi\rangle$, completely quantifies also the mixing of $\rho_A$ (and $\rho_B$), it could well happen that local measurements, or bilocal on the two parties, optimal for the determination of the mixing, are as informative as the global ones with respect to
entanglement. In fact, in reducing $|\psi\rangle\langle\psi|$ to $\rho_A\otimes\rho_B$ only the relative phase $\alpha$ is lost, the dependence on directions $\hat{a}$ and $\hat{b}$ and on the entanglement $b$ is preserved. We have found the optimal global and local measurement of $b$. The results obtained following the two strategies are the same, as we will discuss in Section \ref{sec:conclusions}, so all the extractable information about the entanglement is preserved under the partial trace operation, and the four classes considered above turn out to be equivalent for entanglement estimation.

\subsection{Effective mixed state}
%
%  effective mixed state
%

 Notice that all the dependence on the measuring strategy (\ref{POVM}) in Eq. (\ref{avKul2}) is contained in the probability $p(k|b)$ of outcome $k$ conditioned to the entanglement of the state being some given $b$,
\begin{equation}
  p(k\vert b)=\int\limits_{S^2}\frac{d\hat{a}}{4\pi}\int\limits_{S^2}
    \frac{d\hat{b}}{4\pi}\int\limits_{S^1}\frac{d\alpha}{2\pi}
    \thinspace\mbox{tr}(M^{(k)}\rho^{\otimes N}) ,
\label{prob}
\end{equation}
where the sum over the rest of parameters reflects the fact that we are only interested in the entanglement. This expression can also be written as
\begin{equation}
  p(k\vert b)=\mbox{tr}(M^{(k)}\rho^{(N)}(b)) ,
\label{prob2}
\end{equation}
where the mixed state $\rho^{(N)}(b)$ is
\begin{equation}
  \rho^{(N)}(b)\equiv\int\limits_{S^2}\frac{d\hat{a}}{4\pi}\int\limits_
    {S^2}\frac{d\hat{b}}{4\pi}\int\limits_{S^1}\frac{d\alpha}{2\pi}
    \thinspace|\psi\rangle\langle\psi|^{\otimes N} .
\label{mixst}
\end{equation}
Eq. (\ref{prob2}) allows for an alternative interpretation to our
problem: a $2N$-qubit mixed state $\rho^{(N)}(b)$ is drawn randomly with prior probability distribution $f(b)= 3b^2$ and we want to determine it by estimating $b$.
%
% p(k|b)
%

We will compute $p(k\vert b)$ in the basis that diagonalizes $\rho^{(N)}(b)$, which will crucially turn out to be independent of $b$. Let us denote by $\lambda_1(b), ..., \lambda_m(b)$ the positive eigenvalues of $\rho^{(N)}(b)$, and with $n_1, ..., n_m$ their multiplicity. From the normalization of (\ref{mixst}) the relation $\sum_{j=1}^mn_j\lambda_j=1$  follows. The sum $n\equiv \sum_j n_j$ of multiplicities of (non-vanishing) eigenvalues equals the dimension of the space which supports $\proj{\psi}^{\otimes N}$. This is the symmetric subspace of ${\mathcal H}_4^{\otimes N}$, and thus \cite{ALP}
\begin{equation}
  n = \frac{(N+2J)!}{N!(2J)!}=\frac{(N+3)(N+2)(N+1)}{6}.
\label{dimension}
\end{equation}

With this notation Eq. (\ref{prob2}) reads
\bea
  p(k\vert b)&=&\lambda_1(b)\sum_{i=1}^{n_1}M_{ii}^{(k)}+\lambda_2(b)
    \sum_{i=n_1+1}^{n_1+n_2}M_{ii}^{(k)}+\ldots \nonumber\\
&+&\lambda_m(b) \sum_{i=n-n_m+1}^nM_{ii}^{(k)}\equiv \sum_{j=1}^m\lambda_j(b)q_j^{(k)}.
\label{prob3}
\eea
By substituting this expression in (\ref{avKul2}) and using the inequality
\cite{TV}
\begin{equation}
  (x_1+x_2)\ln\left(\frac{x_1+x_2}{y_1+y_2}\right)\le x_1\ln\left(
    \frac{x_1}{y_1}\right)+x_2\ln\left(\frac{x_2}{y_2}\right) ,
\label{inequality}
\end{equation}
where $x_i, y_i \geq 0$, along with the fact that the POVM is a resolution of the identity in the symmetric subspace of ${\mathcal H}_4^{\otimes N}$, i.e. $\sum_kq_j^{(k)}=n_j$, it follows that the average gain of information is bounded by
\begin{equation}
  \bar{K}[f_k,f]\le\int dbf(b)\sum_{j=1}^mn_j\lambda_j(b)\ln
    \left(\frac{\lambda_j(b)}{\int dbf(b)\lambda_j(b)}\right).
\label{optgain}
\end{equation}

\subsection{Minimal most informative measuring strategy.}
%
% need for spectral decomposition and plan
%

The bound (\ref{optgain}) can be minimally saturated through a measurement with $m$ outcomes where each $M^{(k)}$ is the $n_k$-dimensional projector over the subspace corresponding to the
eigenvalue $\lambda_k$ of $\rho^{(N)}(b)$, having then $p(k\vert
b)=n_k\lambda_k(b)$. Therefore the construction of the optimal measurement can
be readily performed after the computation of the spectral decomposition of the state
(\ref{mixst}), and this is done for an arbitrary $N$ in the next Section. For a more detailed account of the $N=1,2,3$ cases see Section \ref{sec:examples}, where also the gain of information up to $N=80$ has been computed explicitly.

%
% independence of the figure of merits
%

 Notice also that there are other ways measuring strategies can be evaluated and, consequently, there is not a unique notion of optimality. For instance, in \cite{HOL,MP,DBE,LPT,ALP,VLPT} a guess for the unknown state is made depending on the outcome of the measurement, and then both guessed and unknown state are compared using the fidelity. It can be proved, following Ref. \cite{TV}, that the optimal measurements presented here, the most informative ones, are also optimal if we decide, alternatively, for a fidelity-like figure of merit satisfying some very general conditions \cite{FID}.

\section{Computation of $\rho^{(N)}$}
\label{sec:computation}
%
% Computation of rho^(N)(b)
%

It has been shown that the spectrum of $\rho^{(N)}(b)$ determines the
maximal gain of information about $b$, whereas its eigenprojectors lead to the corresponding measuring strategy. Our next step will be the computation of the spectral decomposition of this effective mixed state.

Let us rewrite the generic two-qubit pure state (\ref{Schmidt}) as
\bea
  |\psi\rangle &=& U_A\otimes U_B \left( c_+\ket{+}_A\otimes\ket{+}_B + c_-\ket{-}_A\otimes\ket{-}_B \right)\nonumber\\
    &\equiv& U_A\otimes U_B\ket{\psi(b)} ,
\label{Schmidt2}
\eea
where $c_{+} \equiv \sqrt{\frac{1+b}{2}}$, $c_{-} \equiv \sqrt{\frac{1-b}{2}}$, the single-qubit pure states $\ket{+}_A$ and $\ket{-}_A$ ($\ket{+}_B$ and $\ket{-}_B$) constitute an orthonormal basis in Alice's (Bob's) part --corresponding to some fixed direction in the Bloch sphere--, $U_A$ and $U_B$ are unitary transformations in each single-qubit space and $|\psi(b)\rangle$ is a reference state.

 The state $\rho^{(N)}(b)$ corresponds then to a Haar integral over the group
$SU(2)\times SU(2)$, since it can be expressed as
\begin{equation}
  \rho^{(N)}(b)=\int_{g\in G} dg \left(D(g)M(b)D(g)^{\dagger}\right)^{\otimes N} ,
\label{mixst2}
\end{equation}
where the index $g$ denotes the elements of the group $G=SU(2)\times SU(2)$,
$D(g)=U_A\otimes U_B$ is a $\frac{1}{2}\times\frac{1}{2}$ irreducible representation (irrep) of this group and $M(b)=|\psi(b)\rangle\langle\psi(b)|$.

A well-known result in group representation theory following from Schur's lemma, the so-called orthogonality lemma, will be useful in the calculation of this integral. Consider a matrix $A^{\alpha\beta}(B)$
given by
\begin{equation}
  A^{\alpha\beta}(B)=\int_{g\in G}dgD^\alpha(g)BD^{\beta\dagger}(g) ,
\label{object}
\end{equation}
where $D^\alpha$ and $D^\beta$ are two unitary irreps of the group $G$. Then,

\vspace{5mm}

\noindent {\bf Lemma 1 (orthogonality lemma):}
\begin{equation}
  A^{\alpha\beta}(B)=a(B)\delta^{\alpha\beta}I ,
\label{Shur}
\end{equation}
so $A^{\alpha\beta}(B)$ is zero if the two representations are inequivalent
and proportional to the identity if the two representations are equivalent.

\vspace{5mm}

In order to benefit from this lemma we identify $B$ with $M(b)^{\otimes N}=\proj{\psi(b)}^{\otimes N}$ and then consider the relevant irreps of $SU(2)\times SU(2)$ borne by the $N$-fold tensor product of the $\frac{1}{2}\times \frac{1}{2}-$irrep of the group. These representations are the support of the state $\ket{\psi(b)}^{\otimes N}$, and our next task is to recognize them.

%
% relevant irreducible representations
%

The state $\ket{\psi(b)}^{\otimes N}$ can be expanded as
\begin{eqnarray}
&&\ket{\psi(b)}^{\otimes N}=c_+^N\ket{\!+\!+\! ...\!+\!+\!}_A \otimes\ket{.}_B 
    \nonumber \\
%----------
&&+c_+^{N-1}c_-\bigg(\!\!
\ket{\!+\!+\! ...\!+\!-\!}_A\otimes\ket{.}_B + \cdots +\ket{\!-\!+\! ... \!+\!+\!}_A\otimes\ket{.}_B \!\!\bigg)\nonumber \\
%----------
&&+c_+^{N-2}c_-^2\bigg(\!\!
\ket{\!\!+\! ...\!+\!--\!}_A\otimes\ket{.}_B + \cdots +\ket{\!-\!-\!+\! ... \!+\!}_A\otimes\ket{.}_B \!\!\bigg)\nonumber \\
%----------
&&+ c_+^{N-3}c_-^3 \bigg( \,\,\,\, \bigg) + \cdots + c_+c_-^{N-1} \bigg( \,\,\,\,\bigg) \nonumber \\
%----------
&&+c_-^N\ket{\!-\!-\! ...\!-\!-\!}_A \otimes\ket{.}_B,
\label{stateN}
\end{eqnarray}
where $\ket{.}_B$ means that we have exactly the same vector in the second subsystem. Notice that in the expression above all the elements of the product basis $\{\ket{u_i}\}$
of the local spaces ${\mathcal H}_2^{\otimes N}$ of Alice's and Bob's $N$ qubits --i.e. $\ket{u_1}=\ket{\!+\!+\! ...\!+\!+\!}, \ket{u_2}=\ket{\!+\!+\! ...\!+\!-\!}, \cdots, \ket{u_{2^N}}=\ket{\!-\!-\! ...\!-\!-\!}$-- appear in the form $\ket{u_i}_A\otimes\ket{u_i}_B$. Notice, in addition, that if we denote by $m_T$ the sum of the third spin component of all spinors in each ket --i.e., for instance $m_T(\ket{+++}) = 3/2$, $m_T(\ket{++-}) = 1/2$, $m_T(\ket{-+-}) = -1/2$, ...--, the terms multiplied by the same combination of the factors $c_+$ and $c_-$ have the same $m_T$ in $A$ and $B$. The state (\ref{stateN}) can thus also be expressed as
\bea
\ket{\psi(b)}^{\otimes N}=&c_+^N&\sum_{i;m_T=\frac{N}{2}}\ket{u_i}_A\otimes\ket{u_i}_B \nonumber\\
 +c_+^{N-1}&c_-&\sum_{i;m_T=\frac{N}{2}-1} \ket{u_i}_A\otimes\ket{u_i}_B + \cdots \nonumber\\
 &+c_-^N& \sum_{i;m_T=-\frac{N}{2}} \ket{u_i}_A\otimes\ket{u_i}_B.
\label{stateN2}
\eea
We move now from the local spin basis $\{\ket{u_i}_A\}$ to the coupled one $\{\ket{v_i}_A\}$ in Alice's $N$ qubits, and we also do the same in Bob's. The following lemma, that can be easily checked, will be useful here.

\vspace{5mm}

{\bf Lemma 2:} Let $\{\ket{e_i}\}$ and $\{\ket{f_i}\}$ be two orthonormal basis in ${\cal C}^l$, related by an orthogonal transformation $O$, so that  $\ket{e_i}=\sum_j O_{ij}\ket{f_j}$, with $O^*=O$, and $O^{-1}=O^{\dagger}$. Then, 
\be
\sum_{i=1}^l \ket{e_i}\otimes\ket{e_i} = \sum_{i=1}^l \ket{f_i}\otimes\ket{f_i}.
\ee 
\vspace{5mm}

Now, notice that the unitary transformation relating the local basis and the coupled one is real (since all the Clebsch-Gordan coefficients are real) and that there is a conservation rule for the total third spin component (i.e. the Clebsch-Gordan
coefficients that couple two states with third component $m_1$ and $m_2$ to a coupled state with third component $m$ are proportional to $\delta_{m,m_1+m_2}$). Then Eq. (\ref{stateN2}) can be reexpressed, using the previous two facts and lemma 2, in the coupled basis as

\bea
\ket{\psi(b)}^{\otimes N}=&c_+^N&\sum_{i;m_T=\frac{N}{2}}\ket{v_i}_A\otimes\ket{v_i}_B \nonumber\\
 +c_+^{N-1}&c_-&\sum_{i;m_T=\frac{N}{2}-1} \ket{v_i}_A\otimes\ket{v_i}_B + \cdots \nonumber\\
 &+c_-^N& \sum_{i;m_T=-\frac{N}{2}} \ket{v_i}_A\otimes\ket{v_i}_B.
\label{stateN3}
\eea
(see the examples in next Section for more details). We note that the symmetry between the terms in $A$ and in $B$ allows us to derive (\ref{stateN3}) from (\ref{stateN2}).
 
Let us now have a closer look into Eq. (\ref{stateN3}). The term with coefficient $c_+^N$ corresponds simply to the state with a total spin $j$ maximal in both Alice's and Bob's subsystem (i.e., $j_A = j_B = \frac{N}{2}$) and also maximal third spin component $m$, namely $m_A=m_B= \frac{N}{2}$. We can thus write, with the notation $\ket{^{j_A} m_A}_A\otimes\ket{^{j_B} m_B}_B$, $\ket{v_1}\equiv \ket{v_1}_A\otimes\ket{v_1}_B=\ket{^{\frac{N}{2}}\frac{N}{2} }_A\otimes\ket{^{\frac{N}{2}} \frac{N}{2}}_B$. This state belongs to a $\frac{N}{2}\otimes\frac{N}{2}$-irrep of the group $SU(2)\times SU(2)$. The coefficient $c_+^{N-1}c_-$ corresponds to all states with $m_A=m_B= \frac{N}{2}-1$. Apart from $\ket{v_2}\equiv\ket{^{\frac{N}{2}}\frac{N}{2}\!-\!1 }_A\otimes\ket{^{\frac{N}{2}}\frac{N}{2}\!-\!1}_B$, which again belongs to the previous $\frac{N}{2}\otimes\frac{N}{2}$-irrep, the remaining $N\!-\!1$ kets, $\ket{v_3}\cdots\ket{v_{N+1}}$ have $j_A=j_B=\frac{N}{2}\!-\!1$, and thus belong to $N\!-\!1$ different (but equivalent) $(\frac{N}{2}\!-\!1)\otimes(\frac{N}{2}\!-\!1)$-irreps of the group. But since only the linear combination $\ket{v_3} + \cdots + \ket{v_{N\!+\!1}}$ appears, the relevant irrep is just the symmetric combination of the latter $N\!-\!1$ ones, which we will denote by $\{(\frac{N}{2}\!-\!1)\otimes(\frac{N}{2}\!-\!1)\}_{sym}$, and which no longer decomposes as the product of two irreps of $SU(2)$. The same applies for $(\frac{N}{2}\!-\!2)\otimes(\frac{N}{2}\!-\!2)$-irreps and so on.

 Thus, the space which supports the initial state can be decomposed in terms
of irreps of $SU(2)\times SU(2)$ as
\bea
 \frac{N}{2}\!\otimes\!\frac{N}{2}\oplus\bigg\{\!\left(\!\frac{N}{2}\!-\!1\!\right)\!\otimes\left(\frac{N}{2}\!-\!1\right)\bigg\}_{sym}\oplus\ldots\nonumber\\
\oplus  \bigg\{\frac{N mod\thinspace 2}{2}\otimes\frac{N mod\thinspace 2}{2} \bigg\}_{sym} ,
\label{space}
\eea
where $N mod\thinspace 2$ is equal to one for odd $N$ and equal to zero
for even $N$. It can be checked that this result agrees
dimensionally with formula (\ref{dimension}).

The decomposition shown above in terms of the relevant irreps of
the group $SU(2)\times SU(2)$ together with the orthogonality lemma can be used to solve the integral in (\ref{mixst2}). As we have argued, when plugging (\ref{stateN3}) into (\ref{mixst2}) the cross terms corresponding to inequivalent representations --such as $\ket{v_1}(\bra{v_3}+....+\bra{v_{N\!+\!1}})$-- vanish as we integrate, while the terms within the same representation --such as $\proj{v_1}$-- lead to a contribution proportional to the
identity in the subspace associated with the representation. So the state
$\rho^{(N)}(b)$ is equal to
\bea
  \rho^{(N)}(b)=\lambda_1(b)I_{\frac{N}{2}\otimes\frac{N}{2}}&+& \lambda_2(b)I_{\{\left(\frac{N}{2}\!-\!1\right)\otimes\left( \frac{N}{2}\!-\!1\right)\}_{sym}}+\ldots\nonumber \\
&+&\lambda_m(b)I_{\{\frac{N mod\thinspace 2}{2}\otimes\frac{N mod\thinspace 2}{2}\}_{sym}} .
\label{spectrum}
\eea
This is the spectral decomposition we are looking for, where $\{\lambda_j\}$ are the entanglement dependent eigenvalues of $\rho^{(N)}(b)$, the trace of the identities giving the corresponding multiplicities $\{n_j\}$. It is important to notice that, as it was mentioned before, the eigenspaces are independent of $b$.

The calculation of $n_j\lambda_j$ can now be readily performed from Eq. (\ref{stateN3}) by computing the trace of the projection of $\ket{\psi(b)}^N$ into each relevant irrep. The determination of the spectrum of $\rho^{(N)}(b)$ completes, as we have shown, the
construction of the optimal measurement for the estimation of the
entanglement. In the next section some examples are studied in order to
clarify the implementation of the procedure.

\section{Some examples: the $N=1,2,3$ cases and beyond.}
\label{sec:examples}

In this section we will apply the procedure described above to obtain
the optimal estimation of $b$ when one, two and three identical copies
of the initial state are at our disposal.

\subsection{$N=1$}

The simplest case, $N=1$, is now straightforward. The state written as in
(\ref{Schmidt2}) belongs to the $\frac{1}{2}\otimes\frac{1}{2}$
irrep of $SU(2)\times SU(2)$. From (\ref{mixst2}) we have,
using the orthogonality lemma as in (\ref{spectrum}),
\begin{equation}
  \rho^{(1)}(b)=\int dg D(g)M(b)D(g)^{\dagger}=\lambda_1(b)I.
\end{equation}
The eigenvalue $\lambda_1(b)=\frac{1}{4}$ is obtained by taking the trace in the expression above. The probability $p(k\vert b)$ (see (\ref{prob2})) is independent of $b$, so that $p(k)=p(k\vert b)$ and the average Kullback information (\ref{avKul2}) vanishes.

 Consequently, no information whatsoever can be obtained about the entanglement of a completely unknown pure state if only one copy is at our disposal.

\subsection{$N=2$}

For the $N=2$ case the initial state has the form, from (\ref{stateN})
or (\ref{stateN2}),
\bea
  \ket{\psi(b)}^{\otimes 2}&=&c_+^2\ket{\!+\!+\!}_A\otimes\ket{.}_B\nonumber\\
+c_+&c_-&\left(\ket{\!+\!-\!}_A\otimes\ket{.}_B+\ket{\!-\!+\!}_A\otimes\ket{.}_B\right)\nonumber\\
&& +c_-^2\ket{\!--\!}_A\otimes\ket{.}_B,
\eea
Now, using lemma 2 and the conservation law mentioned before for the Clebsch-Gordan coefficients (cf. Eq. (\ref{stateN3})), we can rewrite the state as
\bea
  \ket{\psi(b)}^{\otimes 2}&=&c_+^2\ket{^1 1}_A\otimes\ket{.}_B\nonumber\\
+c_+&c_-&\left(\ket{^1 0}_A\otimes\ket{.}_B+\ket{^0 0}_A\otimes\ket{.}_B\right)\nonumber\\
&& +c_-^2\ket{^1 -\!\!1}_A\otimes\ket{.}_B,
\label{irrep2}
\eea
where for each party the coupled basis is related to the local one by means of an orthogonal transformation, as usual, 
\bea
\ket{^1 1} = \ket{++}; ~~~\ket{^1 -\!\!1} = \ket{--};\nonumber\\
\ket{^1 0} = \frac{1}{\sqrt{2}}\bigg(\ket{+-}+\ket{-+}\bigg);\nonumber\\
\ket{^0 0} = \frac{1}{\sqrt{2}}\bigg(\ket{+-}-\ket{-+}\bigg).
\eea
The state $|\psi(b)\rangle^{\otimes2}$ in (\ref{irrep2}) is supported then in the $1\otimes1$- and the $0\otimes0$- irreps of $SU(2)\times SU(2)$, and now the application of lemma 1 gives for $\rho^{(2)}(b)$
\begin{equation}
  \rho^{(2)}(b)=\lambda_1(b)I_{1\otimes 1}+\lambda_2(b)I_{0\otimes
       0} .
\end{equation}
We just need to pick up the contributions of (\ref{irrep2}) to each irrep, that is the trace of the corresponding projections, to find that
\begin{eqnarray}
  &&n_1\lambda_1(b)=\left(c_+^4+c_+^2c_-^2+c_-^4\right)=
    \thinspace\frac{3+b^2}{4}\nonumber \\
  &&n_2\lambda_2(b)=c_+^2c_-^2=\frac{1-b^2}{4}.
\label{spectrum2}
\end{eqnarray}

 The optimal measurement (see Eq. (\ref{optgain})) then consists of two projectors onto the $1\otimes 1$- and $0\otimes 0$-irreps of $SU(2)\otimes SU(2)$, with probabilities $p(1\vert b)=n_1\lambda_1(b)=\frac{3+b^2}{4}$ and $p(2\vert
b)=n_2\lambda_2(b)=\frac{1-b^2}{4}$, and from them $p(1)=\frac{9}{10}$ and
$p(2)=\frac{1}{10}$. Finally the gain of information can be computed using
(\ref{avKul2}) and it gives $\bar{K}=0.0375$ bits.

\subsection{$N=3$}

The last case we want to discuss is $N=3$. Starting now from
(\ref{stateN3}) we have
\bea
 && \ket{\psi(b)}^{\otimes 3}=c_+^3\ket{^\frac{3}{2}\frac{3}{2}}_A\otimes\ket{.}_B\nonumber\\
&&+c_+^2c_-\left( \ket{^\frac{3}{2}\frac{1}{2}}_A\otimes\ket{.}_B+\ket{^{\frac{1}{2}}\frac{1}{2}}_A\otimes\ket{.}_B+\ket{^{\frac{1}{2}'}\frac{1}{2}}_A\otimes\ket{.}_B \right) \nonumber \\
  &&+c_+c_-^2\left( \ket{^\frac{3}{2}\!-\!\!\frac{1}{2}}_A\!\!\otimes\ket{.}_B+\ket{^{\frac{1}{2}}\!-\!\!\frac{1}{2}}_A\!\!\otimes\ket{.}_B+\ket{^{\frac{1}{2}'}\!-\!\!\frac{1}{2}}_A\!\!\otimes\ket{.}_B \right) \nonumber \\
&&+c_-^3 \ket{^\frac{3}{2}\!-\!\frac{3}{2}}_A\otimes\ket{.}_B,
\label{example3}
\eea
we observe that only contributions to the $\frac{3}{2}\otimes\frac{3}{2}$- and to two different $\frac{1}{2}\otimes\frac{1}{2}$- irreps of $SU(2)\times SU(2)$ appear. Notice, in addition, that since in this expansion the contributions to $\frac{1}{2}\otimes\frac{1}{2}$ and to $\frac{1}{2}'\otimes\frac{1}{2}'$ only appear in a symmetric linear combination (i.e. $\ket{^{\frac{1}{2}}\frac{1}{2}}_A\otimes\ket{.}_B+\ket{^{\frac{1}{2}'}\frac{1}{2}}_A\otimes\ket{.}_B$ and $\ket{^{\frac{1}{2}}\!-\!\!\frac{1}{2}}_A\!\!\otimes\ket{.}_B+\ket{^{\frac{1}{2}'}\!-\!\!\frac{1}{2}}_A\!\!\otimes\ket{.}_B$), the relevant irreps is precisely a symmetric combination of the two latter ones, $\{\frac{1}{2}\otimes\frac{1}{2}\}_{sym}$. The orthogonality lemma gives now
\begin{equation}
  \rho^{(3)}(b)=\lambda_1(b)I_{\frac{3}{2}\otimes\frac{3}{2}}+
    \lambda_2(b)I_{\{\frac{1}{2}\otimes\frac{1}{2}\}_{sym}} .
\end{equation}
Finally, by collecting the traces of each projection of (\ref{example3}) onto each irreps we obtain
\begin{eqnarray}
  &&n_1\lambda_1(b)=\left(c_+^6+c_+^4c_-^2+c_+^2c_-^4+c_-^6\right)
    =\thinspace\frac{1+b^2}{2} \nonumber \\
  &&n_2\lambda_2(b)=\thinspace 2\left(c_+^4c_-^2+c_+^2c_-^4\right)
    =\thinspace\frac{1-b^2}{2},
\label{spectrum3}
\end{eqnarray}
and thus the optimal measurement is composed by a 16-dimensional and a 4-dimensional projectors into the two irreps shown above, the corresponding probabilities being $p(1\vert b)=\frac{1+b^2}{2}$ and $p(2\vert b)=\frac{1-b^2}{2}$. From them $p(1)=\frac{4}{5}$ and $p(2)=\frac{1}{5}$, and the gain of information is of 0.084 bits.

\subsection{$N>3$}

 We have applied the same, general procedure to obtain the gain of information up to $N=80$, as reported in Table \ref{infvsN} and Figure \ref{avinfvsN}. 
We observe a logarithmic asymptotic dependence of the gain of information on the number $N$ of available copies of $\ket{\psi}$, which reads
\be
\bar{K} \approx 0.44 \log_2 N
\ee  
bits of information on $b$.

\section{Optimal estimation of mixing}
\label{sec:mix}

%
% optimal estimation of mixing
%

So far we have considered the most general measurement involving
the whole space $\left({\mathcal H}_2\otimes{\mathcal H}_2\right)^{\otimes N}$ of $N$ copies of a two-qubit pure state. Now we are going to study optimal {\em local} measurements for the estimation of its entanglement. Alice will perform a collective measurement over the $N$ copies of the state $\rho_A$ in Eq. (\ref{redmatr}) at her disposal in order to estimate the parameter $b$. Consequently, we are also studying optimal strategies for estimating the degree of mixing of a single-qubit mixed state, when $N$ copies are available.

 In order to study the latter with more generality we will consider a generic prior distribution $f(b)$ for the degree of mixing while keeping an isotropic distribution in the Bloch vector direction $\hat{a}$ of the unknown mixed state, with
\be
  \int\limits_{S^2}\frac{d\hat{a}}{4\pi}\int_0^1 db\thinspace f(b)=1.
\label{norm2}
\ee
A general measurement on the local composite system supporting the state $\rho_A^{\otimes N}$ consists of a resolution of the identity in the corresponding Hilbert space ${\mathcal H}_2^{\otimes N}$ by means of positive operators $M^{(k)}$. The gain of information is
as in (\ref{avKul2}), where now
\begin{equation}
  p(k\vert b)=tr\left(M^{(k)}\rho_A^{(N)}(b)\right) ,
\label{locprob}
\end{equation}
so that we need to compute the effective mixed state
\begin{equation}
  \rho_A^{(N)}(b)\equiv\int_{g\in G} dg\left(D(g)\rho_A(b)D(g)^{\dagger}
    \right)^{\otimes N} ,
\label{locmixst}
\end{equation}
where the integral is performed over the group $G = SU(2)$ and a single copy of the mixed state
\begin{equation}
  \rho_A=U_A\thinspace\rho_A(b)U_A^{\dagger}
\label{redmatr2}
\end{equation}
has been expressed, as before, in terms of a reference state $\rho_A(b)\equiv\left(c_+^2\proj{+}+c_-^2\proj{-}\right)$ and a unitary transformation $U_A$. The procedure to be followed is analogous to the previous one, the spectral decomposition of the state (\ref{locmixst}) allowing us to build the optimal measurement.

The density matrix $\rho_A(b)^{\otimes N}$ can be written --by using a straightforward modification of lemma 2 and the mentioned properties of the Clebsh-Gordan coefficients-- in terms of the coupled basis $\{\ket{v_i}_A\}$ as 
\bea
  \rho_A(b)^{\otimes N}=&c_+^{2N}&\sum_{i;m_T=\frac{N}{2}}\proj{v_i}_A \nonumber\\
+c_+^{2(N-1)}&c_-^2&\sum_{i;m_T=\frac{N}{2}-1}\proj{v_i}_A+ \ldots \nonumber\\
&+c_-^{2N}&\sum_{i;m_T=-\frac{N}{2}}\proj{v_i}_A.
\label{locstN}
\eea
Notice that the important role played before by the symmetry between the kets in A and B (cf. Eq. (\ref{stateN3})) is now played by the symmetry between the terms in the bra and in the ket. However we see that now there are no cross terms between inequivalent irreps of $SU(2)$, and that equivalent irreps, such as the $N-1$ copies of the $(\frac{N}{2}\!-\!\!1)-$irrep, obtain equal but independent contributions. The space ${\mathcal H}_2^{\otimes N}$, decomposed in terms of irreps of $SU(2)$, is (see also Refs. \cite{VLPT} and \cite{CEM})
\bea
{\mathcal H}_2^{\otimes N} = \frac{N}{2}\oplus\left(\frac{N}{2}\!-\!\!1\right)\oplus\ldots\oplus\left (\frac{N}{2}\!-\!\!1\right)\oplus\ldots\nonumber \\
\oplus\frac{N mod\thinspace 2}{2} \oplus\ldots\oplus\frac{N mod\thinspace 2}{2} .
\label{locspace}
\eea

The spectral decomposition of $\rho_A^{(N)}(b)$ is determined by application of the orthogonality lemma. Since equivalent irreps receive always the same contributions in the decomposition (\ref{locstN}), the corresponding eigenvalues are equal, so that (\ref{locmixst}) reads
\bea
  \rho_A^{(N)}(b)=\lambda_1^L(b)I_{\frac{N}{2}}+\lambda_2^L(b)
    \left(I_{\frac{N}{2}\!-\!\!1}+\ldots+I_{\frac{N}{2}\!-\!\!1}\right)+\ldots \nonumber \\
+ \lambda_m^L(b)\left(I_{\frac{N mod\thinspace 2}{2}}+\ldots+ I_{\frac{N mod\thinspace 2}{2}}\right) .
\label{locspectrum}
\eea
This is, of course, simply what remains from Eq. (\ref{spectrum}) when Bob's subsystem is traced out, and we have included the whole derivation only for completeness.

 Eqs. (\ref{prob3}-\ref{optgain}) still hold and therefore the optimal measurement for the degree of mixing $b$ corresponds, for any isotropic distribution, to projections onto each of the subspaces associated with the eigenvalues $\{\lambda_k^L\}$. The gain of information is then given by the right hand side of Eq. (\ref{optgain}). Notice that both the number of outcomes and the corresponding probabilities $p(k\vert b)=n_k^L\lambda_k^L(b)$ are equal to the ones obtained before for entanglement estimation. In particular, it follows that there is no way to learn about the degree of mixture of an unknown mixed state if only one copy is available.

\section{Discussion and conclusions}

\label{sec:conclusions}

%
% conclusion
%

 We have presented in this work an optimal strategy for the estimation of the entanglement of two-qubit pure states, when $N$ copies are available. Such optimal measurement is also minimal, in the sense that it consists of the minimum number of outcomes, namely $N/2+1 ~((N+1)/2)$ outcomes for the even (odd) $N$-copy case. Most of the corresponding projectors are of dimension greater than one, and of course any further decomposition of them can be used in principle to obtain, simultaneously, some additional information about other properties of the unknown state, although our optimal POVM is not compatible with projecting onto states of the form $\ket{\psi_i}^{\otimes N}$ as optimal POVM for state determination do \cite{MP,DBE,LPT,ALP}, and they are thus less powerful for that purpose. 

An interesting particular case is when the initial state is a product one, i.e. $b=1$. It can be seen that in this situation we have only the outcome corresponding to the space of maximum spin, since $n_1\lambda_1(1)=1$. Therefore if the outcome $k$, with $k>1$, is obtained we can assure that the state is entangled.

In the previous Section we have also been concerned with the optimal estimation of the degree of mixing. Our optimal measurement, again minimal, can be used, for instance, to quantify the degree of purity of states created by a preparation device whose polarization direction we ignore. Our strategy is actually complementary to the one aiming at revealing optimally the direction of polarization of the state \cite{HOL}. As a matter of fact, the optimal POVM we have obtained is just a coarse graining of the one obtained in \cite{VLPT} for optimal estimation of mixed states, which turns out to reach also the optimal standards of direction estimation obtained in \cite{HOL}. Consequently, direction and modulus of the Bloch vector of an unknown mixed state can be optimally estimated simultaneously. Notice that this is not a frequent situation. If, instead, we would like to estimate the $x, y$ and $z$ components of the Bloch vector independently, we would have obtained incompatible optimal strategies (consider e.g. the $N=1$ case, where an optimal measurement for the component of the Bloch vector along direction $\hat{n}$ consists of a two outcome measurement projecting on that direction).

 Finally, we can argue that {\em bilocal} measurements, either {\em uncorrelated} or {\em classically correlated}, do not imply any improvement on the simpler, {\em local} ones for entanglement estimation. Once we get an outcome from Alice's local measurement we can compute Bob's effective state, and it is clear from Eq. (\ref{spectrum}) that his outcome will be the same as Alice's, so that no extra information on $b$ will be obtained. We have also seen that the optimal global measurement on $\ket{\psi}^{\otimes N}$ is perfectly mimicked by a local one on $\rho_A^{\otimes N}$ (or $\rho_B^{\otimes N}$), so that actually all four classes of measurements considered in Section \ref{subsec:types_measurement} are equivalent. In fact, with hindsight, one can understand this result: local measurements are performed on the reduced density matrix, which is obtained by a partial trace over the other subsystem. This operation erases the information contained in the parameters $\alpha$ and $\hat{b}$ of Eq. (\ref{Schmidt}). On the other hand the global measurement can be interpreted as being performed on the effective density matrix of Eq. (\ref{mixst}), where the same parameters have been integrated over. This operation erases the information contained in them too.

It would be challenging to address the same question for bipartite mixed states, and for systems shared by more than two parties. Notice that in none of these cases optimal estimation of the non-local parameters would be possible by means of local (or even uncorrelated bilocal) measuring strategies. This is the case for mixed states because any given reduced density matrix $\rho_A$ may correspond to infinitely many mixed states $\rho$, with different degrees of entanglement, so that not even in the limit $N\rightarrow \infty$ can the entanglement of $\rho$ be properly inferred from $\rho_A^{\otimes N}$. The mere existence of hidden non-local parameters \cite{Kem}, that is of entanglement parameters that are erased during the partial trace operation, also prevents uncorrelated local strategies from being optimal for estimation of pure-state tripartite entanglement.

To conclude, two-qubit pure-state entanglement, a quantum non-local property, can be optimally estimated by means of local, but collective, measurements.

\section*{Acknowledgments}
\begin{figure}
\begin{center}
 \epsffile{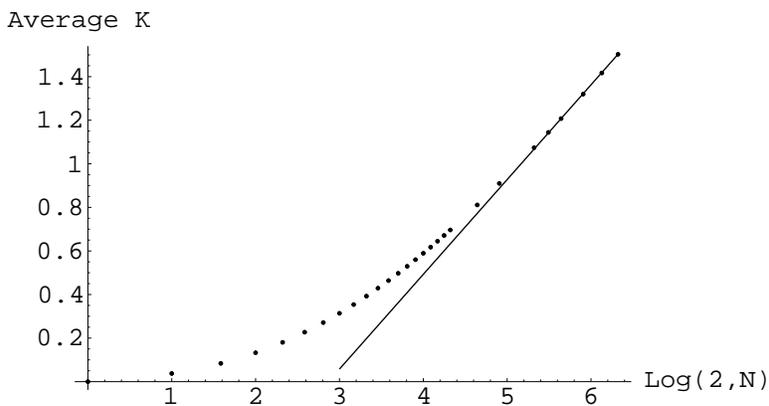}
\caption{
Average gain of information $\bar{K}$ about $b$ given $N$ copies of the state $|\psi\rangle$. The points represent the results obtained by the described optimal measurement, while the line shows the asymptotic behavior.}
\label{avinfvsN}
\end{center}
\end{figure}

We thank Susana Huelga for reactivating our interest in this problem and for interesting discussions, and J. I. Latorre for helping us with the computation of the values of the Figure \ref{avinfvsN}. G.V. acknowledges a CIRIT grant 1997FI-00068 PG. A.A. acknowledges a grant from MEC. Financial support from CICYT contract AEN98-0431 and  CIRIT contract 1998SGR-00026 are also aknowledged.
This work was partially elaborated during the `Complexity, Computation and the Physics of Information' workshop of the Isaac Newton Institute, July
1999. The authors thank the Institute and the European Science
Foundation for support during this period.

\newpage

\begin{table}
\begin{tabular}{||c|lc||}
$~~N~~$ & $~~\bar{K}~~$& \\ 
%\hline
  1 & 0 &\\
  2 & 0.03751 &\\
  3 & 0.08397 &\\
  4 & 0.13259 &\\
  5 & 0.18059 &\\
 10 & 0.39245 &\\
 20 & 0.69639 &\\
 40 & 1.07422 &\\
 60 & 1.32005 &\\ 
 80 & 1.50261 &\\ 
%hline
\end{tabular}
\caption{Average gain of information $\bar{K}$ about $b$ given $N$ 
copies of the state $|\psi\rangle$.}
\label{infvsN}
\end{table}

\bigskip

\begin{references}
\bibitem{FID}  More specifically, the most informative measurements presented in this work are also optimal with respect to a fidelity-guided scheme if the quality of the guesses is evaluated through any {\em concave} fidelity function $F(b-b_k)$ --where $b$ is the unknown parameter and $b_k$ is the guess made after outcome $k$-- that reasonably takes its maximum for $b_k=b$, i.e. $F((x+x')/2)\geq (F(x)+F(x'))/2$ and $F(0) \geq F(x\in [\!-\!1,1])$.
\end{references}
\end{document}